\documentclass[aps,preprint,tightenlines,showpacs,nofootinbib]{revtex4}
\usepackage{amsmath}
\usepackage{amssymb}
\usepackage{epsfig}

\begin{document}

\title{Meson-baryon threshold effects in the light-quark baryon spectrum}
\author{P. Gonz\'alez}
\affiliation{Departamento de F\' \i sica Te\'orica e IFIC, Universidad de Valencia -
CSIC, E-46100 Burjassot, Valencia, Spain}
\author{J. Vijande}
\affiliation{Departamento de F\' \i sica Te\'orica e IFIC, Universidad de Valencia -
CSIC, E-46100 Burjassot, Valencia, Spain}
\affiliation{Departamento de F\'\i sica Fundamental, Universidad de Salamanca, E-37008
Salamanca, Spain}
\author{A. Valcarce}
\affiliation{Departamento de F\'\i sica Fundamental, Universidad de Salamanca, E-37008
Salamanca, Spain}
\date{\today}

\begin{abstract}
We argue that selected $S$ wave meson-baryon channels may play a key role to
match poor baryon mass predictions from quark models with data. The
identification of these channels with effective inelastic channels in data
analysis allows to derive a prescription which could improve the extraction
and identification of baryon resonances.
\end{abstract}

\pacs{14.20.-c,14.20.Gk}
\maketitle

\section{Introduction}

In the Particle Data Group (PDG) book~\cite{PDG06} the light-quark ($u$ and $%
d$) baryon spectrum is composed of forty resonances rated from one ($\ast )$
to four $(\ast \ast \ast \ast )$ stars. The PDG average--mass region below
1950 MeV contains mostly four--star (well established) resonances, fourteen
out of twenty three, the same being true for the $\Lambda$ strange sector,
eight out of eleven. This makes this mass region the most suitable for
testing any spectroscopic quark model. From the pioneering Isgur and Karl's
non-relativistic quark model in the late 70's~\cite{IK78} more refined
spectroscopic quark models for baryons, based on two-body interactions, have
been developed~\cite{SBG85,Cap86,Sta91,Wu94,Glo96,Cap00,VijXX}. We will
refer to them as two-body quark models and we shall denote them generically
as $3q^{2b}$. As an overall result the masses of the fourteen four-star
resonances, most times with the exception of $N_{P_{11}}(1440)$ (see
comments below), are rather well predicted ($\lesssim 100$ MeV difference
with the PDG average value) by these models. Regarding the five three-star
(likely to certain existence) resonances, the situation is much less
favorable since the masses of two of them, $\Delta _{P_{33}}(1600)$ and $%
\Delta _{D_{35}}(1930),$ are generally overpredicted, up to 250 MeV above
the PDG average value. Let us note that a similar discrepancy is observed
for $\Delta _{S_{31}}(1900)(\ast \ast )$ and $\Delta _{D_{33}}(1940)(\ast )$
($\gtrsim 100$ MeV difference with the PDG average value) which can be
related to $\Delta _{D_{35}}(1930)$ as we shall show, and for $\Delta
_{P_{31}}(1750)(\ast )$ (up to 200 MeV above the PDG average). In the
strange $\Lambda $ sector an outstanding overpredicted (by $80-150$ MeV)
state is the $\Lambda _{S_{01}}(1405)(\ast \ast \ast \ast )$. Henceforth we
shall call anomalies these significantly overpredicted mass resonances.

In this article we carry out a general analysis of the anomalies: we
identify them and we propose a plausible physical mechanism to give
correctly account of their masses. To accomplish this task we shall first
examine in detail in Sec.~\ref{secII} the $3q^{2b}$ mass predictions and
advance, through arguments of universality and consistency, the plausible
role played by the coupling of three-quark components ($3q$) to relevant
meson-baryon ($mB$) channels. In Sec.~\ref{secIII} our qualitative
considerations will be put on a more sound basis through a simplified model
calculation. The successful description attained will drive us to prescribe
in Sec.~\ref{secIV} the implementation of these relevant $mB$ channels in
data analysis to improve the extraction of the anomalies. In Sec.~\ref{secV}
we revise alternative partial descriptions from existing quark models
incorporating three-body interactions. Finally, in Sec.~\ref{secVI} we
summarize our main findings.

\section{Two-body quark-model predictions and meson-baryon threshold effects}

\label{secII}

\subsection{Large-energy-step anomalies}

As explained next, most anomalies may be assigned either to a large radial
energy excitation or to a $3q^{2b}$ configuration with large mass induced by
quark Pauli blocking. We shall refer to them as large-energy-step anomalies.

\subsubsection{Radial excitations: $\Delta _{P_{33}}(1600),N_{P_{11}}(1440)$}

The $\Delta _{P_{33}}(1600)$ is the first positive parity excitation of $%
\Delta_{P_{33}}(1232)\equiv \Delta$. The large mass for $\Delta
_{P_{33}}(1600)$ predicted by $3q^{2b}$ can be understood making use of an
harmonic oscillator approximation, with $SU(6)\times O(3)$ symmetry ($%
SU(6)\supset SU(3)_{\mathrm{Flavor}}\times SU(2)_{\mathrm{Spin}};$ for
non-strange quarks the flavor is specified by the isospin $I$). Then the $%
\Delta _{P_{33}}(1600)$ may be assigned to the $(56,L^{P}=0^{+})^{S=3/2}$
configuration in the $\mathcal{N}=2$ band (we shall obviate $I=3/2$ for $%
\Delta$ and $I=1/2$ when referring to nucleon). The band number $\mathcal{N}$
can be expressed as $\mathcal{N}=(2n_{\rho }+\ell _{\rho })+$ $(2n_{\lambda
}+\ell _{\lambda })$ where $\rho $ and $\lambda $ refer to the two Jacobi
coordinates in a three-quark baryon and $\ell_\rho$ and $\ell_\lambda$ to
the corresponding orbital angular momenta. The total orbital angular
momentum of the system is given by $\vec{L}=\vec{\ell _{\rho }}+\vec{\ell
_{\lambda }}$, and the parity $P$ by $P=(-)^{\ell _{\rho }+\ell _{\lambda
}}=(-)^{\mathcal{N}}$. More specifically the $\Delta_{P_{33}}(1600)$ may be
assigned to the first radial excitation of the $\Delta_{P_{33}}(1232)$: $%
(n_{\rho },n_{\lambda })=(1,0)$ or $(0,1)$ and $\ell _{\rho }=0=\ell
_{\lambda }$. From the harmonic oscillator energy, $E=(\mathcal{N}+3)\hbar
\omega $ being $\omega $ the angular frequency, the first radial excitation,
involving an $\mathcal{N}$ increase of two units, is higher in energy than
the first orbital one, ($\ell _{\rho },\ell _{\lambda })=(1,0)$ or $(0,1)$
and $n_{\rho }=0=n_{\lambda }$, for which $\mathcal{N}$ increases only one
unit. However this contradicts data since the $\Delta _{P_{33}}(1600)$ has
lower mass than $\Delta _{D_{33}}(1700)$ or $\Delta _{S_{31}}(1620),$ the
lowest negative parity excitations. This inversion problem, equivalent to
the mass overprediction for $\Delta_{P_{33}}(1600)$, appears also for the
Roper resonance, $N_{P_{11}}(1440)$, lower in mass than $N_{D_{13}}(1520)$
and $N_{S_{11}}(1535)$. Actually the solution of the Roper inversion has
motivated many \textit{ad hoc} quark model studies. Being our goal to get as
much as possible a general understanding of the anomalies we shall include,
in parallel to $\Delta _{P_{33}}(1600)$, the Roper resonance in our list.

Let us add that the conventional interpretation of $N_{P_{11}}(1440)$ and $%
\Delta _{P_{33}}(1600)$ as radial excitations we have assumed does not
preclude other configuration assignments. In $N_{P_{11}}(1440)$ there is
mixing with the orbital excitation $(70,0^{+})_{\mathcal{N}=2}^{S=1/2}$
which could even be dynamically dominant. In $\Delta _{P_{33}}(1600)$ an
alternative interpretation in terms of orbital excitations is also feasible
as we shall show later on.

\subsubsection{Quark Pauli blocking induced states: $\Delta
_{D_{35}}(1930),\Delta _{P_{31}}(1750)$}

Regarding the $\Delta _{D_{35}}(1930),$ say the lowest $\Delta (5/2^{-})$
energy state, it can be assigned to the $(56,1^{-})^{S=3/2}$ configuration~%
\cite{Cut76,Dal77}. Although the expression of $\mathcal{N}$ ($\mathcal{N}%
\geq L)$ may suggest $\mathcal{N}=1$, this energy band is forbidden since
being a completely symmetric state in isospin, $I=3/2$, and spin, $S=3/2$,
the spatial part should also be completely symmetric whereas $\mathcal{N}=1$
only admits spatial states of mixed symmetry. Instead $\mathcal{N}=3$
according to parity with an $\mathcal{N}$ increase of two units, hence its
predicted large mass. So quark Pauli blocking makes the system acquire two
units of excitation, this time in the form ($\ell _{\rho },\ell _{\lambda
})=(1,2)$ or $(2,1)$, instead of ($\ell _{\rho },\ell _{\lambda })=(1,0)$ or 
$(0,1)$.

An analogous situation occurs for $\Delta _{P_{31}}(1750)$, the lowest $%
\Delta (1/2^{+})$ PDG state. The configuration assigned, $(70,0^{+})^{S=1/2}$%
, cannot combine with $\mathcal{N}=0$ which only admits completely symmetric
spatial states, then $\mathcal{N}=2$ through two units of excitation ($\ell
_{\rho },\ell _{\lambda })=(1,1)$, instead of ($\ell _{\rho },\ell _{\lambda
})=(0,0)$.

\subsection{Meson-baryon threshold channels}

Given the large radial excitation energy and the large mass predicted for
quark Pauli blocking induced states, one may wonder about the possibility
that $4q1\overline{q}$ components may be energetically competitive, despite
the extra quark and antiquark masses. Thus, they could greatly contribute,
altogether with $3q$ components, to the formation of the bound structures%
\footnote{%
Indeed the $\Delta _{D_{35}}(1930)$ was first interpreted as a hybrid state
involving gluonic or nonvalence quark degrees of freedom~\cite{Cut76}
although this interpretation was questioned a few years later~\cite{Bow80}
through a revision of the role played by anharmonic perturbations.}. In
order to examine this possibility at a phenomenological level we look for $%
4q1\overline{q}$ components in the form of inelastic meson-baryon channels
in relative $S$ wave (the lowest energy partial wave) with adequate quantum
numbers to couple to the anomalies and with thresholds close above their PDG
masses. We shall name these components meson-baryon threshold channels or $%
mB $ channels.

\subsubsection{$\Delta _{D_{35}}(1930)$, $\Delta _{D_{33}}(1940)$ and $%
\Delta_{S_{31}}(1900)$}

For $\Delta _{D_{35}}(1930)$ a simple inspection allows us to identify the
following $mB$ channels: $\pi \Delta _{F_{35}}(1905),$ $\omega \Delta$, and $%
\rho \Delta $ with thresholds at 2045 MeV, 2014 MeV, and 2002 MeV,
respectively (let us recall that $3q^{2b}$ mass predictions are $80-250$ MeV
higher than the PDG average 1930 MeV). To discriminate among these channels
we notice that going further with our argumentation we should expect the
presence of $\Delta$ resonances close in mass to $\Delta _{D_{35}}(1930)$,
whenever the same dominant configuration $(56,1^{-})_{\mathcal{N}=3}^{S=3/2}$
and the same relevant thresholds are present. We shall refer to these
resonances as partners. It turns out that $\Delta (1/2^{-})$ and $\Delta
(3/2^{-})$ contain that configuration (in fact it is the only one common to
these two deltas and $\Delta (5/2^{-})$ below 2.2 GeV). Moreover if we
examine the PDG table we find the anomalies $\Delta _{S_{31}}(1900)$ and $%
\Delta _{D_{33}}(1940)$ sharing with $\Delta _{D_{35}}(1930)$ the $\omega
\Delta $ and $\rho \Delta $ as $mB$ channels. This suggests $\Delta
_{S_{31}}(1900)$ and $\Delta _{D_{33}}(1940)$ as partners of $\Delta
_{D_{35}}(1930)$ and $\omega \Delta $ and/or $\rho \Delta $ as the possible
relevant coupling to the binding of the three resonances.

Let us add for the sake of completeness that for $\Delta (1/2^{-})$ and $%
\Delta (3/2^{-})$ the $3q^{2b}$ first radial excitation $(70,1^{-})_{%
\mathcal{N}=3}^{S=1/2}$, at about $2050\pm 50$ MeV, is not far above the
average mass of their anomalies. However we shall justify later on the
assignment of these radial excitations to $\Delta _{S_{31}}(2150)(\ast )$
and to a not yet extracted $\Delta _{D_{33}}$ resonance around the same
energy.

\subsubsection{$\Delta _{P_{31}}(1750)$}

An analogous analysis based on the search of $mB$ channels can be carried
out for $\Delta _{P_{31}}(1750)$ with a $60-200$ MeV mass overprediction
from $3q^{2b}.$ We find $\pi N_{S_{11}}(1650)$ and $\pi \Delta
_{S_{31}}(1620)$ with thresholds at 1790 MeV and 1760 MeV, respectively.
Since both thresholds involve pions with $J^{P}=0^{-}$ there cannot be $%
\Delta$ ($J\neq 1/2$) partners, with $(70,0^{+})_{\mathcal{N}=2}^{S=1/2}$
and the same relevant thresholds. In consequence we have no further
phenomenological indication on which threshold may be relevant.

\subsubsection{$\Delta _{P_{33}}(1600)$ and $N_{P_{11}}(1440)$}

For $\Delta _{P_{33}}(1600)$ with a $3q^{2b}$ mass overprediction of $80-250$
MeV, the $\pi N_{D_{13}}(1520)$ channel (threshold at $1660$ MeV) might
contribute to the binding. Additionally $\sigma \Delta $ with a quite
uncertain $S$ wave threshold due to the large interval accepted for the $%
\sigma$ mass ($400-1200$ MeV) might play some role.

Analogously for $N_{P_{11}}(1440)$, with $3q^{2b}$ mass predictions ranging
from 1410 MeV to 1700 MeV, the $\sigma N$ channel could play a relevant
role. In fact, the explicit consideration of $\sigma N$ has allowed for a
description of $N_{P_{11}}(1440)$ from a coupled meson-baryon channel
calculation~\cite{Kre00}. The $\pi N_{S_{11}}(1535)$ channel, although with
threshold at 1675 MeV quite above the PDG mass, could also have some effect.
In both cases ($J_{\pi }=0=J_{\sigma })$ there are no partners to be
examined.

\subsection{Regular-energy-step anomalies}

Certainly meson-baryon channel coupling effects may be at work for other
resonances not involving either large energy excitation steps or a large
mass induced by quark Pauli blocking.

\subsubsection{$\Lambda _{S_{01}}(1405)$}

As a matter of fact the more generally accepted anomaly is the $\Lambda
_{S_{01}}(1405)$ which has motivated a lot of studies being mostly
interpreted, at the hadron level, as an $S$ wave $N\overline{K}$ quasi-bound
system (in the chiral unitary approximation one of the poles couples mostly
to $N\overline{K}$~\cite{Lambda}). Alternatively, at the quark level, the
identification of the lowest $3q$ negative parity excitation of $\Lambda $
(predicted mass about $1550$ MeV) with $\Lambda _{S_{01}}(1405)$ has been
suggested, the difference in mass being attributed to the mass shift induced
in $3q(\Lambda (1/2^{-}))$ by its strong coupling to the $S$ wave $N%
\overline{K}$ channel (threshold at 1435 MeV)~\cite{Cap86,Pak99}. Very
recently a quantitative calculation along this line within a specific quark
model framework has been performed~\cite{Tak07}. Let us remark that although
these explanations are formulated in terms of different degrees of freedom
(hadrons or quarks) they may be somehow equivalent through the effectiveness
of parameters, cutoffs...

Henceforth we shall assume that $\Lambda _{S_{01}}(1405)$ is a resonance
induced by the coupling of $N\overline{K}$ to the lowest energy $3q^{2b}$
negative parity configuration with strangeness: $(70,1^{-})_{\mathcal{N}%
=1}^{S=1/2}$ and flavor singlet, $I=0$. Since $J^{P}=0^{-}$ for kaons no $%
\Lambda _{S_{01}}(1405)$ partners are expected. In fact the closest $\Lambda 
$ resonance, $\Lambda _{D_{03}}(1520)$, shares the same configuration but
has no coupling to $N\overline{K}.$

\subsubsection{$\Delta _{F_{35}}(\sim 1720)$}

For other light-quark resonances in the energy region under consideration ($%
\leq 1950$ MeV) the inspection of data and $3q^{2b}$ mass predictions makes
us conclude that mass overpredictions are not very significant with one
possible exception. This corresponds to the lowest energy state of $\Delta
(5/2^{+}).$ Since $L\neq 0$ the minimum possible $\mathcal{N}$ value is $%
\mathcal{N}=2$ according to parity. $3q^{2b}$ models predict two states in
the $\mathcal{N}=2$ band with mass ranges $1870-1940$ MeV and $1930-2030$
MeV what seems to be in correspondence with the first and second PDG states $%
\Delta _{F_{35}}(1915)(\ast \ast \ast \ast )$ and $\Delta
_{F_{35}}(2000)(\ast \ast ).$ However the $\Delta _{F_{35}}(2000)$ is
bizarre since its average mass is obtained from three different data
analyses, two of them~\cite{Man92,Vra00} reporting a mass about $1720$ $(\pm
60)$ MeV and the other~\cite{Cut80} giving a quite different value of $%
2200\pm 125$ MeV. Then by considering two differentiated resonances the $%
\Delta _{F_{35}}(\sim 1720)$ would be a clear candidate for an anomaly.
Remarkably there is a $mB$ channel, the $\pi N_{D_{15}}(1675)$ with
threshold at $1815$ MeV, which could contribute to the binding of this
resonance at such low energy. Note additionally that the $3q^{2b}$
configuration corresponding to $L=L_{\min }=1$ ($\implies S=3/2)$ and $%
\mathcal{N}=2$ is forbidden in this case due to its antisymmetric orbital
character. This Pauli blocking does not imply though an increase in $%
\mathcal{N}$ but an orbital reordering of the quarks to an available $L=2$, $%
\mathcal{N}=2$ configuration, hence its regular-energy-step character.

Therefore we will tentatively identify the $\Delta _{F_{35}}(\sim 1720)$ as
the lowest energy state of $\Delta (5/2^{+})$ and interpret it as another
meson-baryon, $\pi N_{D_{15}}(1675)$, induced resonance. For its $3q^{2b}$
configuration assignment there are two options: i) $(70,2^{+})_{\mathcal{N}%
=2}^{S=1/2}$ and ii) $(56,2^{+})_{\mathcal{N}=2}^{S=3/2}$.

If we opt for i) this configuration is also present in $\Delta (3/2^{+})$
where it could couple to $\pi N_{D_{13}}(1700)$ with threshold at $1840$
MeV. It turns out that $N_{D_{13}}(1700)$ is almost degenerate to $%
N_{D_{15}}(1675)$ sharing the same dominant configuration $(70,1^{-})_{%
\mathcal{N}=1}^{S=3/2}$. Then we should expect a companion resonance of $%
\Delta _{F_{35}}(\sim 1720)$ in $\Delta (3/2^{+})$ at about the same energy.
By revisiting the PDG book we find that $\Delta _{P_{33}}(1600)$ is assigned
a mass around $1700$ MeV in Refs.~\cite{Man92,Vra00} in agreement with our
expectation. Therefore our proposal of a distinctive $\Delta _{F_{35}}(\sim
1720)$ should be complemented with the consideration of the current $\Delta
_{P_{33}}(1600)$ as a superposition of a resonance, companion of $\Delta
_{F_{35}}(\sim 1720),$ and of the $3q^{2b}$ first radial excitation
configuration which could also be affected by some $mB$ channel.
Complementarily configuration ii) which is present in $\Delta(5/2^+)$, $%
\Delta(3/2^+)$, and $\Delta(1/2^+)$, would be assigned to $%
\Delta_{F_{35}}(1915)$, $\Delta _{P_{33}}(1920)$ and $\Delta_{P_{31}}(1910)$%
. The almost degenerate mass of these resonances seems to support this
correspondence.

If instead we opt for ii) the same conclusion about the structure of $\Delta
_{P_{33}}(1600)$ would be obtained since ii) also appears in $\Delta
(3/2^{+})$. Besides, as mentioned, ii) is also present in $\Delta (1/2^{+})$
where it could couple to $\pi N_{S_{11}}(1650)$ with threshold at 1790 MeV
(note that $N_{S_{11}}(1650)$ is almost degenerate and shares configuration
with $N_{D_{15}}(1675)$). Then we should expect another companion resonance
of $\Delta _{F_{35}}(\sim 1720)$ in $\Delta (1/2^{+}).$ We could identify
this companion as the $\Delta _{P_{31}}(1750)$. In consequence we should
conclude that the $3q^{2b}$ quark Pauli blocking induced configuration
previously considered for $\Delta _{P_{31}}(1750)$, i.e., $(70,0^{+})_{%
\mathcal{N}=2}^{S=1/2}$, should be instead assigned to the next $\Delta
(1/2^{+})$ resonance $\Delta _{P_{31}}(1910)$ (note that in such a case no
coupling to any relevant $mB$ channel would be needed for this
configuration). Complementarily configuration i) would give account in this
case of $\Delta _{F_{35}}(1915)$ and $\Delta _{P_{33}}(1920)$. Unfortunately
being in any case a resonance induced through pions, $J^{P}=0^{-}$, there
are no $\Delta $ partners which could help to decide in favor of one of the
options.

\begin{figure}[tbp]
\caption{Mass predictions for the anomalies from Ref.~\protect\cite{Cap86}
(dashed lines) as compared to the experimental mass intervals detailed in
Table~\protect\ref{t2} (boxes). N.C. means non-cataloged resonance.}
\label{fig1}\epsfig{file=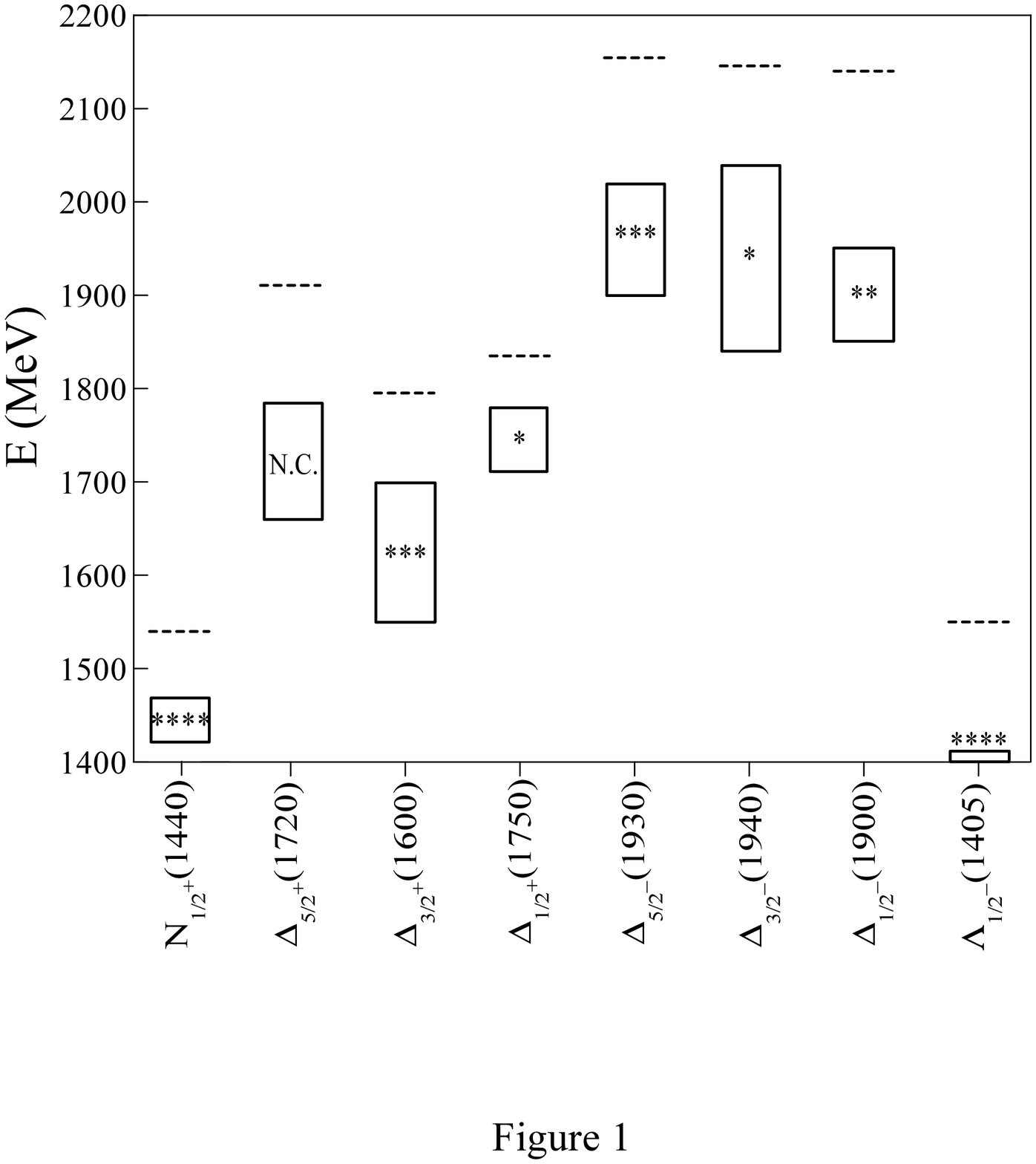,width=14cm}
\end{figure}

To finish this section we represent in Fig.~\ref{fig1} the $3q^{2b}$ mass
predictions based on Ref.~\cite{Cap86} (the numerical values will be given
in Table~\ref{t2}) as compared to the experimental mass intervals for the
anomalies.

\section{Naive model calculation}
\label{secIII}

To go beyond the qualitative analysis of the anomalies we have carried out
some dynamical input is required. In the last years there has been an
important progress in the development of dynamical coupled-channel (DCC)
models of $\pi N$ scattering in the resonance region below $2$ GeV \cite%
{Lee07,Bru07,Spe03}. These models introduce bare baryon states to represent
the quark core components of the resonances. Such components can be
identified with constituent quark model predictions. The resonance, $R$
associated with a bare baryon state is induced by effective vertex
interactions $R\rightarrow mB$ and $R\rightarrow \pi \pi N$. In practice the
masses of the bare states are parameters of the model which are determined
by fitting data (other parameters as effective coupling constants and form
factors are fixed, as much as possible, within reasonable ranges). The
most recent fit to $\pi N$ elastic scattering data (not including $\pi
N\rightarrow \pi \pi N)$ \cite{Bru07} indicates clearly that bare masses are
higher than the PDG's resonance positions. This suggests these
coupled-channel schemes as the appropriate frameworks for a thorough study
of the anomalies. However this is a formidable task out of the scope of this
article which has a rather exploratory character. Instead we shall perform a
simplified quark model calculation along the lines followed in the meson
case to evaluate $2q2\bar{q}$ effects~\cite{Bro04}.

We shall consider a system of one confined channel, the $3q^{2b}$, in
interaction with one free-channel, a meson-baryon threshold channel $mB$,
with a hamiltonian matrix: 
\begin{equation}
\lbrack H]\simeq \left( 
\begin{array}{cc}
M_{m}+M_{B} & a \\ 
a^{\ast } & M_{3q^{2b}}%
\end{array}%
\right) 
\end{equation}%
where $M_{3q^{2b}}$ stands for the mass of the $3q^{2b}$ state, $M_{m}$ and $%
M_{B}$ for the masses of the meson and baryon respectively and $a$ for a
fitting parameter giving account of the interaction ($a$ could correspond
for instance to a $^{3}P_{0}$ transition hamiltonian matrix element). The
effect of the interaction on the masses is easily obtained by
diagonalization. The corresponding eigenvalues are 
\begin{equation}
M_{\pm }=\left( \frac{M_{3q^{2b}}+(M_{m}+M_{B})}{2}\right) \pm \sqrt{\left( 
\frac{M_{3q^{2b}}-(M_{m}+M_{B})}{2}\right) ^{2}+\left\vert a\right\vert ^{2}}
\end{equation}%
where $M_{-}$ is smaller that $(M_{m}+M_{B})$ and $M_{+}$ is bigger than $%
M_{3q^{2b}}$.

It is noteworthy the correspondence between this simplified model and a
truncated DCC model calculation. So $M_{3q^{2b}}$ represents a bare
resonance mass, $M_{m}$ and $M_{B}$ meson and baryon masses which have
been approximated by the experimental values, and $a$ a constant giving
account of the bare resonance$-mB$ effective coupling.

In order to proceed to calculate the eigenvalues we have to choose a
particular $3q^{2b}$ model and establish a criterion for the choice of the $%
mB$ channel for each anomaly. We shall use as $M_{3q^{2b}}$ the values
calculated in Ref.~\cite{Cap86}. Since only the energy band and not the
detailed configuration corresponding to each value has been published an
educated guess has been done. As $mB$ we shall take for granted the $N%
\overline{K}$ channel for $\Lambda _{S_{01}}(1405)$. For $\Delta
_{D_{35}}(1930),$ $\Delta _{D_{33}}(1940)$ and $\Delta _{S_{31}}(1900)$ we
shall select $\rho \Delta $ (equivalently we could have preferred the almost
degenerate $\omega \Delta )$ as suggested by our phenomenological analysis.
For the same reason $\pi N_{D_{15}}(1675)$ will be employed for $\Delta
_{F_{35}}(\sim 1720).$ For $\Delta _{P_{31}}(1750)$ we shall use $\pi
N_{S_{11}}(1650)$ since this coupling is favored at least in one of the two
possible configuration assignments previously discussed.

Regarding $N_{P_{11}}(1440)$ and $\Delta _{P_{33}}(1600)$ the situation is
rather intricate due to the alternative interpretations (radial and/or
orbital excitations) available. From the particular $3q^{2b}$ model we use,
the value $M_{3q^{2b}}=1540$ MeV corresponding to a dominant first radial
excitation of $N$, can be unambiguously assigned to $N_{P_{11}}(1440).$ Then
if we use the nominal PDG average value for the $\sigma $ mass ($600$ MeV)
the $\sigma N$ threshold is about the same energy than $M_{3q^{2b}}$ and
closer to the PDG mass of the Roper than $\pi N_{S_{11}}(1535)$ with
threshold at 1675 MeV. This suggests the selection of the energetically more
competitive $\sigma N$ channel as the possible relevant one. We should keep
in mind though that the selection could be different favoring $\pi
N_{S_{11}}(1535)$ for other choices of the $\sigma $ mass and the $3q^{2b}$
model. Actually given the large mass and width of the $\sigma $ it is also
possible that both $mB$ channels may be contributing to the binding.
Consequently our selection should be considered as an effective one getting
an insight into the relevant meson-baryon threshold effects.

Concerning the $\Delta _{P_{33}}(1600)$ it could be just the companion of $%
\Delta _{F_{35}}(\sim 1720)$ with the $3q^{2b}$ first radial excitation of $%
\Delta $ being hidden in its large width. However, although with different
values of $I$ and $S,$ the first radial excitations of $\Delta $ and $N$
share the same $SU(6)\times O(3)$ configuration. On the other hand the $B$
components of the respective $mB$ channels ($N_{D_{13}}(1520)$ \textit{vs.} $%
\pi N_{S_{11}}(1535)$ and $\Delta $ \textit{vs.} $N)$ also share $%
SU(6)\times O(3)$ configuration. Then it seems natural to assume that
meson-baryon threshold effects may be also acting on the first radial
excitation of the $\Delta$. Moreover the low PDG average mass of $\Delta
_{P_{33}}(1600)$ as compared to $1720$ MeV, the approximated mass of the
companion resonance mentioned above, seems to reinforce this idea. By using
again the PDG nominal $\sigma$ mass we realize that only the $\pi
N_{D_{13}}(1520)$ channel has now a threshold (at 1660 MeV) below the $%
3q^{2b}$ first radial excitation at 1790 MeV. This suggests the selection of 
$\pi N_{D_{13}}(1520)$ as the relevant $mB$ channel (the same caution and
comments as in the Roper case should be applied here). 
\begin{table}[t]
\begin{tabular}{|c|c|}
\hline
PDG average & Meson-baryon channels \\ \hline
$\Delta_{P_{33}}(1600)(***)$ & $\pi\,N_{P_{13}}(1520)$ \\ \hline
$N_{P_{11}}(1440)(****)$ & $\sigma\,N$ \\ \hline\hline
$\Delta_{D_{35}}(1930)(***)$ &  \\ 
$\Delta_{D_{33}}(1940)(*)$ & $\rho\,\Delta$ \\ 
$\Delta_{S_{31}}(1900)(*)$ &  \\ \hline
$\Delta_{P_{31}}(1750)(*)$ & $\pi\,N_{S_{11}}(1650)$ \\ \hline\hline
$\Delta_{F_{35}}(\approx1720)(\mathrm{N.C.})$ & $\pi\,N_{D_{15}}(1675)$ \\ 
\hline
\end{tabular}%
\caption{Light-quarks PDG resonances from Ref.~\protect\cite{PDG06} (N.C.
means non-cataloged) representing spectroscopic anomalies and corresponding
selected meson-baryon threshold channels.}
\label{t1}
\end{table}

In Table~\ref{t1} we list all the light-quark baryon anomalies and the $mB$
channels plausibly contributing to their bindings according to our
discussion (for $\Delta _{P_{33}}(1600)$ we list only the one corresponding
to the radial excitation; for $\Delta_{P_{31}}(1750)$ the listed $mB$
channel refers only to one of the options commented above). In Fig.~\ref%
{fig2} we represent the selected energy thresholds, $M_{m}+M_{B}$ (values in
Table~\ref{t2}) as compared to the experimental mass interval for the
anomalies. 
\begin{table}[b]
\begin{tabular}{|c|cc|cc|c|c|}
\hline
PDG Resonance & $mB$ threshold & Prob. & $3q^{2b}$ & Prob. & $M_-$ & 
Experiment \\ \hline\hline
$\Delta_{P_{33}}(1600)(***)$ & [$\pi\,N_{P_{13}}(1520)$] (1660) & 81.1\% & 
1795 & 18.9\% & 1619 & 1550--1700 \\ \hline
$N_{P_{11}}(1440)(****)$ & [$\sigma\,N$] (1540) & 50.0\% & 1540 & 50.0\% & 
1455 & 1420--1470 \\ \hline
$\Delta_{D_{35}}(1930)(***)$ &  & 83.4\% & 2155 & 16.6\% & 1964 & 1900--2020
\\ 
$\Delta_{D_{33}}(1940)(*)$ & [$\rho\,\Delta]$ (2002) & 82.2\% & 2145 & 17.8\%
& 1962 & 1840--2040$^\dagger$ \\ 
$\Delta_{S_{31}}(1900)(*)$ &  & 81.5\% & 2140 & 18.5\% & 1961 & 1850--1950
\\ \hline
$\Delta_{P_{31}}(1750)(*)$ & [$\pi\,N_{S_{11}}(1650)$] (1790) & 62.8\% & 1835
& 37.2\% & 1725 & 1710--1780 \\ \hline
$\Delta_{F_{35}}(\approx 1720)$(N.C.) & [$\pi\,N_{D_{15}}(1675)$] (1815) & 
74.4\% & 1910 & 25.6\% & 1765 & 1660--1785$^{\dagger \dagger}$ \\ \hline
$\Lambda_{S_{01}}(1405)(****)$ & [$\bar K\,N$] (1434) & 78.2\% & 1550 & 
21.8\% & 1389 & 1400--1410 \\ \hline
\end{tabular}%
\caption{Predicted masses, $M_-$, for the anomalies as compared to
experimental data from Ref.~\protect\cite{PDG06}, Ref.~\protect\cite{Cut80}
(indicated by the superindex $\dagger$), and Ref.~\protect\cite{Vra00}
(indicated by the superindex $\dagger\dagger$). Two-body quark-model masses (%
$3q^{2b}$) are taken from Ref.~\protect\cite{Cap86}. Probabilities (Prob.)
for meson-baryon and $3q$ components are also shown. All masses are in MeV.}
\label{t2}
\end{table}

\subsection{$M_{-}$ resonances}

Although the value of $\left\vert a\right\vert $ might vary depending on the
configurations involved in each $(mB)-3q$ coupling we shall use for the sake
of simplicity the same value in all cases. The $M_{-}$ results for $%
\left\vert a\right\vert =85$ MeV are numerically detailed in Table~\ref{t2}
where the values for $M_{3q^{2b}}$ and for ($M_{m}+M_{B})$ in the chosen $mB$
channel as well as their probabilities to give $M_{-}$ are also displayed.
As can be checked the improvement of the description with respect to $%
3q^{2b} $ is astonishing. All the predicted $M_{-}$ masses lye very close to
the PDG average masses for the anomalies. In Fig.~\ref{fig3} the $M_{-}$
values for $\left\vert a\right\vert =85$ MeV are drawn\ as compared to the
experimental mass intervals.

We interpret these results as providing strong quantitative support to our
former qualitative description of the anomalies. Regarding their nature a
look at the probabilities reveal they are mostly meson-baryon states.
Actually a meson-baryon probability greater or equal than $50\%$ can serve
as a criterion to identify an anomaly. Nonetheless the coupling to the $3q$
component is essential to lower their masses making them more stable against
decay into $m+B$. 

Furthermore the resulting probabilities could be used, at least in the cases
where $M_{m}+M_{B}$ is very close above the PDG average mass, to make a
quantitative estimation of the effective coupling constant of the physical
state to states of the continuum \cite{Han07}.

\begin{figure}[tbp]
\caption{Selected energy thresholds (solid lines) as compared to the
experimental mass intervals for the anomalies detailed in Table~\protect\ref%
{t2} (boxes). N.C. means non-cataloged resonance.}
\label{fig2}\epsfig{file=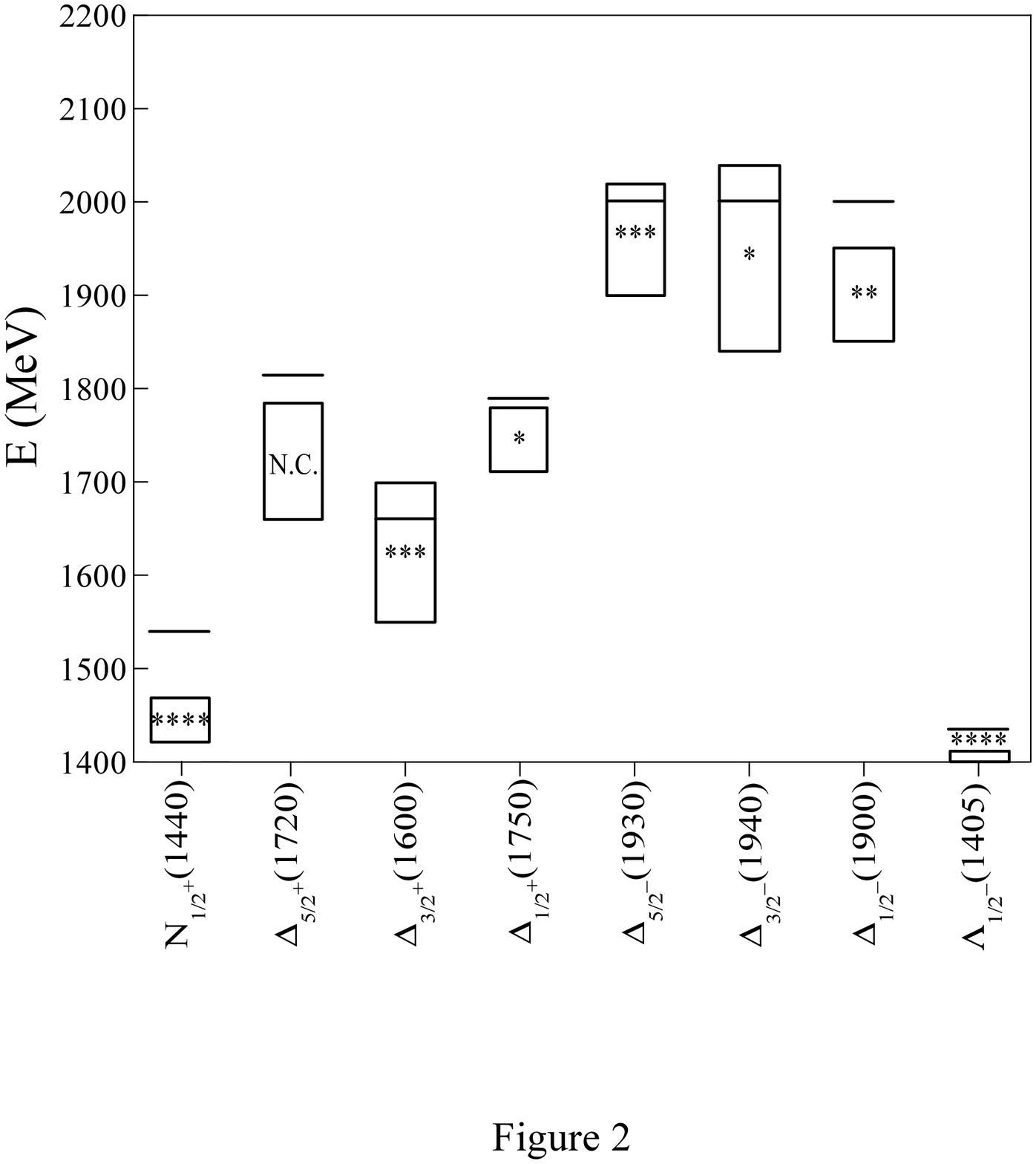,width=14cm}
\end{figure}

\begin{figure}[tbp]
\caption{Predicted masses for the anomalies (dashed lines) as compared to
the experimental mass intervals detailed in Table~\protect\ref{t2} (boxes).
N.C. means non-cataloged resonance.}
\label{fig3}\epsfig{file=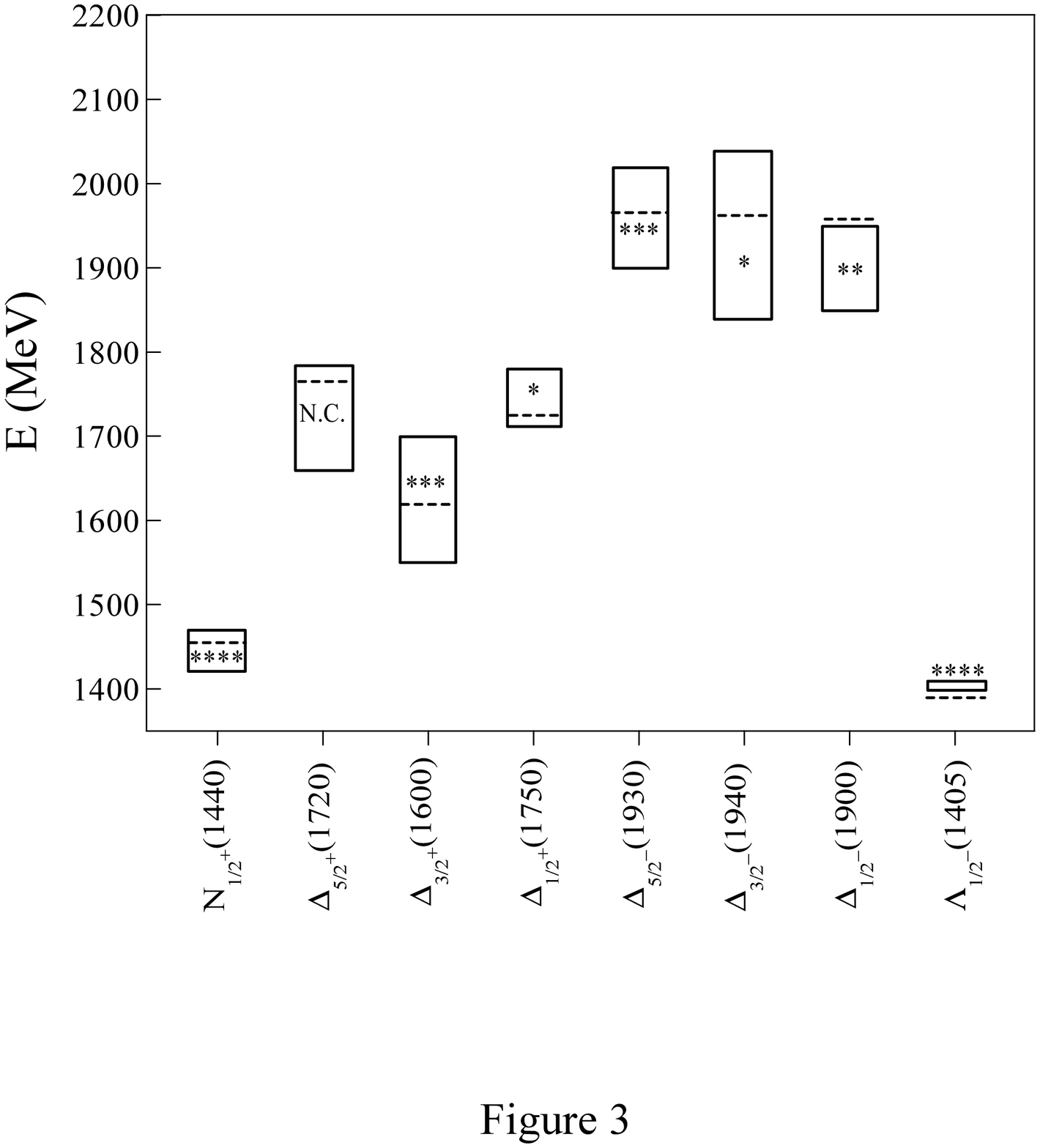,width=14cm}
\end{figure}

It should be emphasized that similar results could be obtained for any other
spectroscopic $3q^{2b}$ model through a fine tuning of the value of $%
\left\vert a\right\vert $ (note that the small value of $\left\vert
a\right\vert $ as compared to $M_{3q^{2b}}$ and $(M_{m}+M_{B})$ provides an 
\textit{a posteriori} validation of our method)$.$ This comes from the
expression of the eigenvalues where it is clear that even for $\left\vert
a\right\vert =0$ one gets $M_{-}=M_{m}+M_{B}$ which according to our $mB$
choice is much closer to the PDG mass of the anomaly than $M_{3q^{2b}}$, see
Fig.~\ref{fig2}. This means that concerning the mass of the anomalies the
coupling of meson-baryon to $3q$ components may play the role of a general 
\emph{healing mechanism} for spectroscopic models.

\subsection{$M_{+}$ resonances}

The values of $M_{+},$ the second solutions, as well as the next $3q^{2b}$
mass predictions taken from Ref.~\cite{Cap86} are shown in Table~\ref{t3} as
compared to the PDG average masses and widths of the anomalies and to the
mass of the next PDG equally labeled resonances. The $M_{+}$ states, with
masses above $M_{3q^{2b}}$, are dominantly $3q$ components (with
complementary $(mB)-3q$ probabilities with respect to $M_{-})$ and may decay
into $m+B.$ 
\begin{table}[tbp]
\begin{tabular}{|c|c|c|c|c|}
\hline
PDG anomaly & $\Gamma$(MeV) & $M_+$ & Next $3q^{2b}$ & Next PDG resonance \\ 
\hline
$N_{P_{11}}(1440)(****)$ & $\sim$ 300 & 1625 & 1770 & $N_{P_{11}}(1710)(***)$
\\ \hline
$\Delta_{P_{33}}(1600)(***)$ & $\sim 350$ & 1836 & 1915 & $%
\Delta_{P_{33}}(1920)(***)$ \\ \hline
$\Delta_{D_{35}}(1930)(***)$ & $\sim 360$ & 2193 & 2165 & $%
\Delta_{P_{35}}(2350)(*)$ \\ 
$\Delta_{D_{33}}(1940)(*)$ & $\sim 200$ & 2185 & 2080 &  \\ 
$\Delta_{S_{31}}(1900)(*)$ & $\sim 200$ & 2180 & 2035 & $%
\Delta_{S_{31}}(2150)(*)$ \\ \hline
$\Delta_{P_{31}}(1750)(*)$ & $\sim 300$ & 1900 & 1875 & $%
\Delta_{P_{31}}(1910)(****) $ \\ \hline
$\Delta_{F_{35}}(\approx1720)(\mathrm{N.C.})$ & $\sim 140$ & 1960 & 1990 & $%
\Delta_{F_{35}}(1905)(****)$ \\ \hline
$\Lambda_{S_{01}}(1405)(****)$ & $50$ & 1595 & 1615 & $%
\Lambda_{S_{01}}(1670)(****)$ \\ \hline
\end{tabular}%
\caption{Predicted masses, $M_{+}$, as compared to the masses of the next $%
3q^{2b}$ states from Ref.~\protect\cite{Cap86} and the next PDG resonances
from Ref.~\protect\cite{PDG06}. The widths of the corresponding anomalies
are also shown. All masses are in MeV.}
\label{t3}
\end{table}
A look at Table~\ref{t3} shows that for $\Delta _{F_{35}}(\sim 1720)$ and $%
\Lambda _{S_{01}}(1405)$ the $M_{+}$ state can be clearly assigned
altogether with the next $3q^{2b}$ state to the next PDG resonances $\Delta
_{F_{35}}(1905)$ and $\Lambda _{S_{01}}(1670)$ respectively. In the first
case both $M_{+}$ and the next $3q^{2b}$ mass are above the next PDG mass
whilst in the second case both lye below it. This is in accord with the
quite general quark model tendency of over (under) prediction for states in
the $\mathcal{N}=2\,\, (1)$ bands.

The same type of assignment can be done for $\Delta _{S_{31}}(1900)$ where
the $M_{+}$ and the next $3q^{2b}$ masses lye respectively close above and
below the next PDG mass average $\Delta _{S_{31}}(2150)$. For the sake of
consistency we should also expect $\Delta _{D_{35}}$ and $\Delta _{D_{33}}$
resonances around 2150 MeV which have not been reported. This may have to do
with the bigger proliferation of $3q^{2b}$ states about that energy (three
for $\Delta (5/2-)$ and four for $\Delta (3/2-)$ against two for $\Delta
(1/2-))$ what may make difficult its experimental disentanglement. Indeed
the large width of $\Delta _{D_{35}}(1930)$ may be also including the effect
of the missed resonance. In any case there is need of further data analysis
to clarify the situation.

For $\Delta _{P_{31}}(1750)$ the $M_{+}$ state could also play some role in
its large width although an assignment to $\Delta _{P_{31}}(1910)$ seems the
most logical.

Finally for $N_{P_{11}}(1440)$ and $\Delta _{P_{33}}(1600)$ the $M_{+}$
states with masses 85 MeV below $N_{P_{11}}(1710)$ and $\Delta
_{P_{33}}(1920)$ respectively may be influencing the large widths of the
anomalies as well as the widths of these next PDG states.

\subsection{Other thresholds}

Let us note that the expression given above for the eigenvalues is symmetric
under the exchange of $M_{3q^{2b}}$ and $(M_{m}+M_{B}).$ Then mass
corrections to $3q^{2b}$ states could alternatively come from meson-baryon
channels above the $3q^{2b}$ mass predictions. However for an anomaly with $%
3q^{2b}$ mass prediction far above the PDG average we do not expect these
contributions to be physically relevant in the sense of having any effect on
its mass. Actually, in our simplified treatment a much larger value of $|a|$
would be required to get a correct mass shift from these thresholds, putting
into question the very validity of the model. On the other hand, $mB$
channels different than the selected ones could have dynamically some
effects through higher partial waves. In the spirit of quark model
calculations we consider the $mB$ channels we have selected (which may not
have a precise experimental correspondence) as effective ones giving account
of the couplings of $3q^{2b}$ states with meson-baryon components.

\section{Resonance Extraction Prescription (REP)}

\label{secIV}

Most light-quark baryon resonances are extracted from data through a
parametrization of $\pi N$ scattering partial waves. This usually refers to
a multichannel scattering matrix including effective inelastic channels. The
consideration of multichannel couplings becomes relevant when an important
channel opens within the width of a resonance. Let us emphasize that this is
so even if the threshold of the channel is above the mass of the resonance.
Actually the consideration of the $\sigma N$ (also named $\epsilon N)$
channel as an effective inelastic channel in some data analyses becomes
relevant for the experimental extraction of the Roper resonance $%
N_{P_{11}}(1440)$, see for instance reference~\cite{Man92}. In parallel the
extraction of $\Delta _{D_{35}}(1930)$ as a distinctive resonance is
associated in some data analysis to the explicit inclusion of a $\rho \Delta 
$ effective inelastic channel~\cite{Man92,Cut80}. Hence a certain
correspondence between efficient inelastic channels in data analyses and our
selected meson-baryon threshold channels considered for the anomalies shows
up for $\Delta _{D_{35}}(1930)$ and $N_{P_{11}}(1440)$. This suggests the
generalization of this correspondence. Therefore we propose the explicit
inclusion in data analyses of the selected meson-baryon threshold channels
in order to make easier the extraction of the anomalies. In this way an
improvement (star-number increase) over the current PDG star-status could
result for all of them as well as consistency among different analysis of
approximately the same set of data might be attained. We shall call REP
(Resonance Extraction Prescription) this proposal.

\section{Alternative explanations: three-body quark-model predictions}

\label{secV}

Our description of the anomalies does not exclude in principle other
possible equivalent effective treatments. As already mentioned alternative
explanations based on poles in a meson-baryon coupled channel approach can
be also found in the literature for $N_{P_{11}}(1440)$ and $\Lambda
_{S_{01}}(1405)$. In this last case even a meson-baryon bound state
interpretation is feasible. From the point of view of quark approaches it is
worthwhile to comment that for light-quark baryons there exist at least two
quark model calculations in the literature beyond $3q^{2b}$, giving a proper
account of the masses of the large-energy-step anomalies. One of these
models~\cite{Des92} incorporates a two-sigma exchange potential apart from a
one-gluon exchange and confinement interactions. The other is a collective
model where baryons appear as vibrations and rotations of a three-quark $Y$%
-shaped string-like configuration~\cite{Bij94}. The energy systematic of
these models is such that the energy step associated to a radial excitation
or a quark Pauli blocking induced configuration gets reduced to
approximately half of its $3q^{2b}$ value in agreement with data. On the
contrary the predictions for regular-energy-step states do not vary
significantly from $3q^{2b}.$ This explains why the $\Lambda _{S_{01}}(1405)$
is out of the systematic (Ref.~\cite{Des92} predicts a mass of 1550 MeV and
Ref.~\cite{Bij94} 1640 MeV) as well as it would be the $\Delta
_{F_{35}}(\sim 1720)$ (1830 MeV in~\cite{Des92} and 1921 MeV in~\cite{Bij94}%
) in case of its confirmation as a distinctive resonance.

Let us notice though that these exceptions could be put in their right
masses by coupling them to the relevant meson-baryon threshold channels. In
this manner a description of similar quality to the one in Sec.~\ref{secIII}
could be reached. Note that the correct large-energy-step model predictions
would not require now relevant meson-baryon threshold channels. Therefore an
alternative spectral description where the large-energy-step anomalies
correspond to $3q$ states is feasible.

\section{Summary and final comments}

\label{secVI}

To summarize we propose that $4q1\overline{q}$ components, in the form of $S$
wave meson-baryon channels which we identify, play an essential role in the
description of the anomalies, say baryon resonances very significantly
overpredicted by three-quark models based on two-body interactions. As a
matter of fact by considering a simplified description of the anomalies as
systems composed of a free meson-baryon channel interacting with a
three-quark confined component we have shown they could correspond mostly to
meson-baryon states but with a non-negligible $3q$ state probability which
makes their masses to be below the meson-baryon threshold. The remarkable
agreement of our results with data in all cases takes us to refine our
definition and propose the dominance of meson-baryon components as the
signature of an anomaly. Relying on the $3q^{2b}$ mass predictions from Ref.~%
\cite{Cap86} the Roper resonance, $N_{P_{11}}(1440)$, might be just in the
limit being a quasianomalous state with a $50\%$ probability of $\sigma N$.
For the other identified anomalies, $\Delta _{P_{33}}(1600),$ $\Delta
_{D_{35}}(1930),$ $\Delta _{D_{33}}(1940)$, $\Delta _{S_{31}}(1900)$, $%
\Delta _{P_{31}}(1750),$ $\Lambda _{S_{01}}(1405)$ and the non-cataloged $%
\Delta _{F_{35}}(\sim 1720),$ the meson-baryon component probability is
magnified.

Though it is probable that these results may vary quantitatively when a more
complete dynamical coupled-channel calculation is carried out we think it is
reasonable not to expect major qualitative changes. Then it is plausible,
given their dominant meson-baryon character, that the Roper resonance and
specially the other \textit{Magnificent Seven} anomalies be dynamically
generated via simplified effective meson-baryon and/or meson-meson-baryon
coupled channel calculations involving only a selected number of channels
and couplings. Indeed this has been shown for $N_{P_{11}}(1440)$ and $%
\Lambda _{S_{01}}(1405),$ the effectiveness of the parameters possibly
taking implicitly into account the nonconsidered three-quark components. In
particular the effective dynamical generation of $\Delta _{F_{35}}(\sim 1720)
$ could be interpreted as given a strong support to our proposal of
considering it a distinctive resonance. Effective meson-baryon coupled
channel studies would be also welcome to clarify the situation for other
anomalies where alternative three-quark descriptions are available. The
information obtained in this manner could be complemented with the one
coming from quark model evaluations of hadronic transition processes in
order to shed some light on the very detailed nature of the anomalies. With
respect to this let us remind that the \textquotedblleft three-body
quark\textquotedblright\ and the \textquotedblleft two-body quark +
meson-baryon\textquotedblright\ wave functions may be rather different.

More complete studies are also needed to extract some conclusion about the
possible anomalous character of some other resonances, apart from the Roper,
in the nucleon sector. Our easy identification of most $\Delta$ anomalies
may have to do with the quite small mixing present in their assigned
anomalous configurations. Actually mixing with non-anomalous ones might play
a role for some nucleon excitations making them not to show up as very
significantly overpredicted mass states. Particularly states in the $%
\mathcal{N}=2$ band: $N_{P_{11}}(1710)(\ast \ast \ast ),$ $%
N_{P_{13}}(1720)(\ast \ast \ast \ast ),$ $N_{F_{15}}(1680)(\ast \ast \ast
\ast )$, would deserve attention.

To finish we should comment on the possible drawbacks of our approach. Our
description, based on a phenomenological analysis and on a healing formula
for mass corrections to quark models predictions, relies on the assumption
of a significant coupling between specific $3q$ states and relevant
meson-baryon channels. No physical mechanism underlying these particular
couplings is detailed. Indeed, our effective treatment might correspond to
different physical mechanisms depending on the anomaly. Note that for $%
\Lambda _{S_{01}}(1405)$, at difference with $\Delta $ cases, a diquark
dominant induced coupling seems to be favored. Moreover, the consideration
that our effective meson-baryon threshold channel might be either replacing
the influence of other couplings (including other meson-baryon partial
waves) or even correcting three-quark models dynamical deficiencies may be
too naive. Besides we should keep in mind that except for $\Delta
_{P_{33}}(1600)$ and $\Delta _{D_{35}}(1930)$ the existence of all the
magnificent anomalies is fair or poorly established rating $(\ast \ast )$ or 
$(\ast )$ in the PDG book or even non-cataloged ($\Delta _{F_{35}}(\sim
1720))$. Actually as established by the PDG editors most high lying states
are questionable. Hence the possibility that some of them do not remain in time is opened.

Keeping in mind these caveats the universality, consistency and simplicity
of our description make us confident that the implemented physical
ingredients will remain essential in further theoretical evaluations. On the
other hand from the experimental point of view the application of our
Resonance Extraction Proposal (REP) of implementing selected meson-baryon
threshold channels in data analyses might add certainty to the existence of
some resonances and at the same time help to reconcile competing and
sometimes not very compatible partial wave analyses. Future work along these
lines would be encouraging.

\section{acknowledgments}
This work has been partially funded by the Spanish Ministerio de
Educaci\'on y Ciencia and EU FEDER under Contract No. FPA2007-65748,
by European Integrated Infrastructure Initiative 506078,
by Junta de Castilla y Le\'{o}n under Contract No. SA016A17, and by the
Spanish Consolider-Ingenio 2010 Program CPAN (CSD2007-00042).

\end{document}